\documentclass{elsart5p}

\usepackage{natbib}

\usepackage{graphicx}

\usepackage{amssymb}

\begin{document}

\begin{frontmatter}



\title{Observational appearances of isolated stellar-mass black hole accretion - theory and observations}


\author[sao]{Beskin, G.}
\author[sai]{Biryukov, A.}
\author[sao]{Karpov, S.}
\author[sao]{Plokhotnichenko, V.}
\author[sao]{Debur, V.}

\address[sao]{Special Astrophysical Observatory of Russian Academy of Sciences,
Russia}
\address[sai]{Sternberg Astronomical Institute of Moscow State University, Moscow, Russia}

\begin{abstract}
General properties of accretion onto isolated stellar mass black holes in the Galaxy are
discussed. An analysis of plasma internal energy growth during the infall is performed.
Adiabatic heating of collisionless accretion flow due to magnetic adiabatic invariant
conservation is $25\%$ more efficient than in the standard non-magnetized gas case.
It is shown that magnetic field line reconnections in discrete current sheets lead to
significant nonthermal electron component formation, which leads to a formation of
a hard (UV, X-ray, up to gamma), highly variable spectral component in addition to
the standard synchrotron optical component first derived by Shvartsman generated by
thermal electrons in the magnetic field of the accretion flow. Properties of accretion
flow emission variability are discussed.
Observation results of two single black hole candidates -
gravitational lens MACHO-1999-BLG-22 and radio-loud x-ray source with featureless
optical spectrum J1942+10 - in optical band with high temporal resolution
are presented and interpreted in the framework of the proposed model.

\end{abstract}


\end{frontmatter}

\section{Accretion onto isolated stellar-mass black holes}
\label{}

Even though more than 60 years have passed since the theoretical prediction of
black holes as an astrophysical objects \citep{oppenheimer} in some sense they
have not been discovered yet.

The features of the black hole -- the compactness (the size
for the mass $M$ is close to the Schwarzschild one $r_g=\frac{2GM}{c^2}$ and
the mass larger than 3$M_{\odot}$ -- are the necessary, but not the sufficient
ones. The deciding property of the black hole is the presence of the event
horizon, instead of with the usual surface. To decide whether it is present in
a given compact and massive object it is necessary to detect and study the
emission generated very near the event horizon.

It is a very complicated task that cannot be easily
performed in x-ray binaries and AGNs due to high accretion rate and so the high 
optical depth of the accreting gas. 
At the same time, single stellar-mass black holes, which accrete interstellar
medium of  low density ($10^{-2} - 1 $cm$^{-3}$), are the ideal case for
detection and study of the event horizon \citep{shvartsman_1971}.

The analysis of existing data on possible black hole masses and
velocities in comparison with the interstellar medium structure shows
 that in the majority of cases in the Galaxy ($>90\%$) the
accretion rate $\dot m= \dot M c^2/L_{edd}$ can't exceed
$10^{-6}-10^{-7}$ \citep{beskin_2005}.

Black holes usually move supersonically (at Mach numbers 2-3). And only in
cold clouds of the interstellar hydrogen ($n\sim10^2-10^5$, $T\sim10^2$) and
at low velocities of motion ($<10$ km/s) the accretion rates may be high and
black hole luminosities may reach $10^{38}-10^{40}$ erg/c.
For typical interstellar medium inhomogeneity 
the captured specific angular momentum is smaller than that on the BH last
stable orbit, and 
the accretion
is always (with the exception of the case of a black hole in a dense
molecular cloud) spherically-symmetric \citep{mii_2005,beskin_2006}.

\begin{figure}
{\centering \resizebox*{1\columnwidth}{!}{\includegraphics[angle=270]{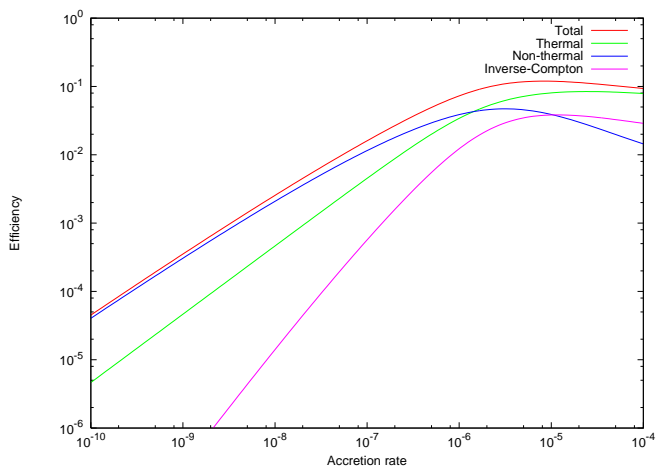} \par}}
  \caption{Efficiencies of the synchrotron emission of thermal and non-thermal
    electron components of the accretion flow.}
  \label{fig_efficiency}
\end{figure}

The accreting plasma is initially collisionless, and it remains so until
the event horizon. The electron-electron and electron-ion free
path $\lambda\sim2.4\cdot10^3T^2n^{-1}$ even at the capture radius is
as high as $\sim10^{12}$ cm. Only the magnetic fields trapped in
plasma (the proton Larmor radius at $r_g$ is 10 cm) make it possible to
consider the problem as a quasi-hydrodynamical one; it is only due to
the magnetic field that the particle's momentum is not conserved, allowing
particles to fall towards the black hole. In addition, the
magnetic field effectively "traps" particles in a "box" of variable
size, which allows us to consider its adiabatic heating during the
fall; a correct treatment of such a process shows that for magnetized
plasma such heating is 25\% more effective than for ideal gas 
due to the conservation of magnetic adiabatic invariant
$I=\frac{3cp_t^2}{2eB}$, 
where $p_t$ is the tangential component of electron momentum\citep{beskin_2005}.
Therefore, the plasma temperature in the
accretion flow grows much faster and electrons become relativistic earlier
--  $R_{rel}\approx6000$ in contrast to $R_{rel}\approx1300$ in 
\citet{bisnovatyi_1974} and
$R_{rel}\approx200$ in \citet{ipser_1982}. The
accretion flow is much hotter, and our estimation of "thermal"
luminosity 
\begin{equation}
L = \eta\dot Mc^2 = 9.6\cdot10^{33} M_{10}^3 n_1^2 (V^2+c_s^2)_{16}^{-3}  \mbox{ erg/s}
\end{equation}
is significantly higher than those of \citet{ipser_1982}
\begin{equation}
L_{\rm IP} = 1.6\cdot10^{32} M_{10}^3 n_1^2 (V^2+c_s^2)_{16}^{-3} \mbox{ erg/s}
\end{equation}
and \citet{bisnovatyi_1974}
\begin{equation}
L_{\rm BKR} = 2\cdot10^{33} M_{10}^3 n_1^2 (V^2+c_s^2)_{16}^{-3} \mbox{ erg/s, }
\end{equation}
while the optical spectral shape is nearly the same. The efficiency of
accretion as a function of accretion rate is shown in Fig.~\ref{fig_efficiency}.

The basis of our analysis is the assumption of the energy
equipartition in the accretion flow (Shvartsman' theorem, see
\citet{shvartsman_1971}).
The straight consequence of this
assumption is the necessity of exceeding magnetic energy dissipation
with the rate
\begin{equation}
\frac{d E}{dV dt} = 4\frac vr
\frac{B^2}{8\pi} - \frac52\frac vr\frac{B^2}{8\pi} =  \frac32\frac{v}{r}\frac{B^2}{8\pi}.
\label{eq_dissipation}
\end{equation}

In the previous accretion models the dissipation goes continuously\citep{bisnovatyi_1997}
  in the turbulent
flow and its mechanism is not examined in details.
We considered conversion of the magnetic
energy in the discrete turbulent current sheets \citep{pustilnik_1997}
 as a mechanism providing such dissipation. At
the same time, various modes of plasma oscillations are generated
(ion-acoustic and Lengmur plasmons mostly), magnetic field lines
reconnect and are ejected with plasma from the current sheet and
electrons are accelerated. The latter effect is very important for the
observational appearances of the whole accretion flow.  The beams of
the accelerated electrons emit its energy due to motion in the
magnetic field and generate an additional (in respect to synchrotron
emission of thermal particles) nonthermal component.

Current sheet spatial scales are usually much smaller than the whole
accretion flow one and, therefore, the fraction of the accelerated particles
is small, and so the total (significantly nonthermal) electron distribution
$f(R, \gamma)$
may be considered as a superposition of the purely thermal one $f_t(R, \gamma)$
for the background
flow particles and purely nonthermal $f_{nt}(R, \gamma)$ for the ones accelerated in the
current sheets (this is the approach known as "hybrid plasma", see \citet{coppi}). Also we assume that for low accretion rates 
the non-adiabatic heating and radiative energy losses do not
change the shape of the thermal component distribution (while changing its mean
energy), then
\begin{equation}
f(R,\gamma) = f_t(R,\gamma)+\zeta f_{nt}(R,\gamma).
\label{eqn_distribution_hybrid}
\end{equation}
Note that this distribution is not normalized, and only its shape has physical
meaning. So, for example, the ratio of nonthermal to thermal electron
densities at some radius $R$ may be expressed as
\begin{equation}
\frac{n_{nt}(R)}{n_{t}(R)} = \frac{\zeta f_{nt}(R)}{f_{t}(R)} \mbox{ ,}
\label{eqn_ratio_nonthermal_thermal}
\end{equation}
where $f_{nt}(R)$ and $f_{t}(R)$ are integrals of the corresponding
distribution functions over the range of $\gamma$.

Thermal particle distribution function may be written as
\begin{equation}
f_t(R,\gamma)=\frac{\sqrt{R}}{2\tau}\left(\frac{\gamma}{\tau}\right)^2\exp
{\left(-\frac{\gamma}{\tau}\right)},
\label{eqn_distribution_thermal}
\end{equation}
where the usual dimensionless expression for temperature $\tau=kT/m_ec^2$ is
used. This gives a Maxwellian local energy distribution and radial density
slope ${\rho\propto R^{-3/2}}$.

To get the temperature profile of the accretion flow let's take into account
that the fraction $\xi\sim0.1$ of the magnetic field energy during its dissipation is converted to the energy
of the accelerated particles. Then, according to (\ref{eq_dissipation}),
the
non-adiabatic heating rate may be expressed as
\begin{equation}
\Phi = (1-\xi)\frac{dE}{dVdt} = (1-\xi)\frac32\frac{v}{r}\frac{B^2}{8\pi}\mbox{ .}
\end{equation}

The main mechanism of radiative losses at low accretion rates is synchrotron
radiation. Its rate may be written as \citep{lightman}
\begin{equation}
\Lambda_{sync} = \frac43\sigma_Tc\overline{\gamma^2}\frac{B^2}{8\pi}n .
\label{eqn_lambda_sync}
\end{equation}

For a Maxwellian distribution
\begin{equation}
\overline{\gamma^2}=12\left(\frac{kT}{m_ec^2}\right)^2=12\tau^{2}.
\end{equation}

For relatively large accretion rates ($\dot m > 10^{-6}$) it is necessary to
take into account the cooling due to inverse Compton effect, which converts the
part of hot electrons energy to the hard emission.

The rate of inverse Compton energy losses is related to the synchrotron one as \citep{lightman}
\begin{equation}
\Lambda_{compton} = \Lambda_{sync}\frac{\epsilon_{rad}}{B^2/8\pi}
\end{equation}

Taking into account non-adiabatic terms, the energy balance equation per particle
may be written for a non-relativistic region of
the flow ($R>R_{rel}$) as 
\begin{equation}
\frac{d}{dr}\frac{\epsilon}{n}=-\frac{5}{4}\frac{\epsilon}{nr}+\frac{1}{v}\frac{\Lambda-\Phi}{n}\mbox{ .}
\end{equation}

For a non-relativistic electron gas $\epsilon=\frac32nkT$, 
so
\begin{equation}
\frac{d\tau}{dR}=-\frac{5}{4}\frac{\tau}{R}-(1-\xi)\frac{\alpha^2}{2}
\frac{m_p}{m_e}R^{-2} ,
\label{temp_eqn_nonrel}
\end{equation}
while for the relativistic one ($\epsilon=3p=3nkT$)
\begin{equation}
\frac{d\tau}{dR}=-\frac58\frac{\tau}{R}-(1-\xi)\frac{\alpha^2}{4}\frac{m_p}{
m_e}R^{-2}+\frac43\frac{m_p}{m_e}\frac{\dot m \tau^2}{R^2},
\label{temp_eqn}
\end{equation}
where $\alpha^2 = \frac13$ (the equipartition condition).

The boundary condition is ${\tau(R_{rel})=1}$.

An analytical solution of this equation in general is very difficult,
but for low accretion rates we may neglect the influence of radiative
losses and get a solution in the form
\begin{equation}
\tau(R)=(1-\xi)\frac{2\alpha^2}{3}\frac{m_p}{m_e}R^{-1}+\left(1-(1-\xi)
\frac{2\alpha^2}{3R_{rel}}\frac{m_p}{m_e}\right)\left(\frac{R_{rel}}{R}
\right)^{5/8}.
\end{equation}
This expression may be substituted into (\ref{eqn_distribution_thermal}) to
get the final expression for the thermal electron distribution.

The energy evolution of single nonthermal electron is described by
\begin{equation}
\frac{d\gamma}{dR} = \frac13\frac{m_p}{m_e}\dot m\frac{\gamma^2}{R^2} -
\frac58\frac{\gamma}{R},
\label{eqn_gamma_evolution}
\end{equation}
where the first term corresponds to synchrotron losses and the second one 
to adiabatic heating.

Analysis of the evolution of electron beams ejected from the current sheets
\citep{beskin_2005}
gives the possibility to compute the $\xi$ and $f_{nt}(R, \gamma)$, i.e. the
fraction of thermal and nonthermal component contributions to the total one. 

\begin{figure}
{\centering \resizebox*{1\columnwidth}{!}{\includegraphics[angle=270]{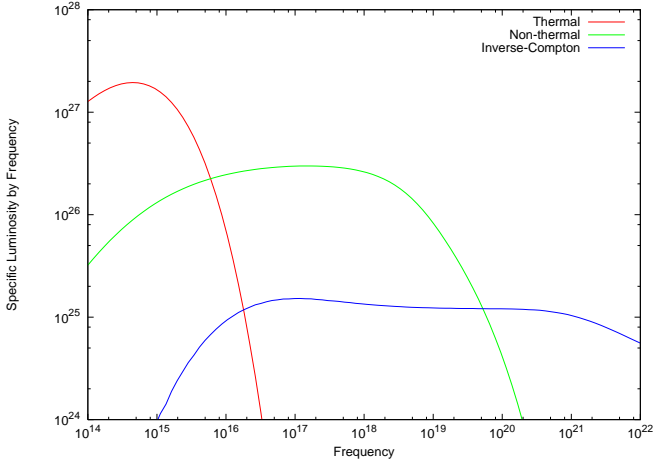} \par}}
  \caption{Decomposition of a single black hole (with the mass $10 M_{\odot}$)
    emission spectrum into thermal and nonthermal parts. The accretion rate is
    $1.4\cdot 10^{10}$ g/s, which corresponds to $\dot m = 10^{-8}$.}
  \label{fig_spectrum}
\end{figure}

\begin{figure}
{\centering \resizebox*{1\columnwidth}{!}{\includegraphics[angle=270]{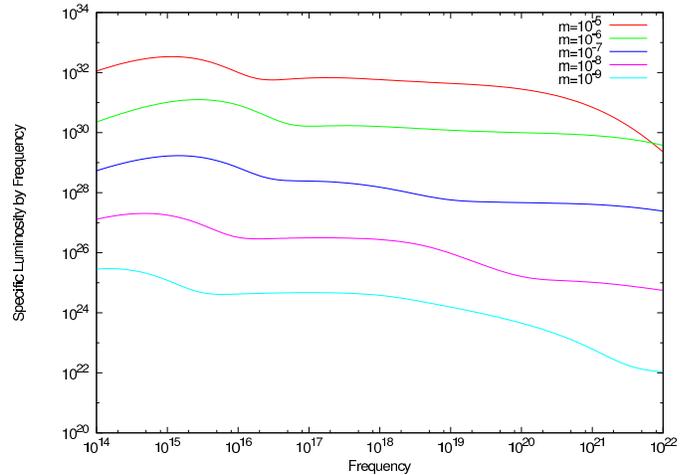} \par}}
  \caption{Spectra of the accretion flow onto the 10 $M_{\odot}$ black hole for the
    various accretion rates.}
  \label{fig_spectra}
\end{figure}

\begin{figure}
  {\centering \resizebox*{1\columnwidth}{!}{\includegraphics[angle=270]{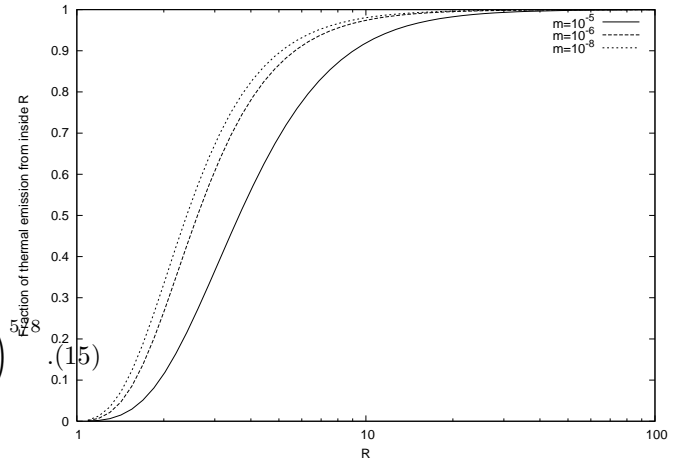}} \par}
  \caption{Fraction of thermal synchrotron emission that comes from inside a given radius $R$}
  \label{fig_luminosity_radial}
\end{figure}

The spectra of various components of the accretion flow emission have been
computed taking into account relativistic effects of emission propagation near
the black hole \citep{lightman,shapiro}. They are shown in
Fig.~\ref{fig_spectra} for various accretion rates. In Fig.~\ref{fig_luminosity_radial}
the integral dependence of the thermal emission of accretion flow on the
distance to the black hole is presented. The large fraction of it is generated
inside the 2$r_g$ sphere, and so carries the information very near the event horizon.

\begin{figure}
{\centering \resizebox*{1\columnwidth}{!}{\includegraphics[angle=0]{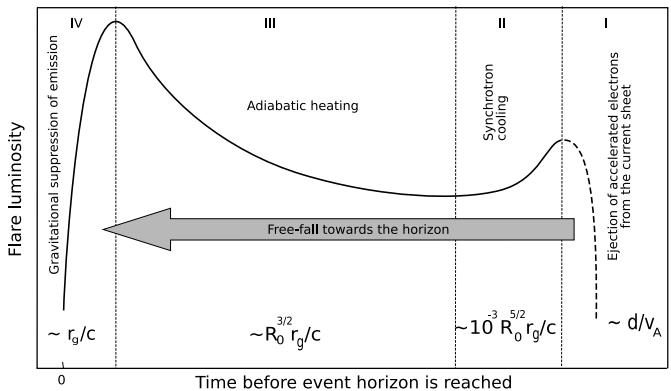} \par}}
  \caption{Internal structure of a flare as a reflection of the electron cloud evolution. 
    The prevailing
    physical mechanisms defining the observed emission are denoted and typical durations of the
    stages are shown.}
  \label{fig_flare_structure}
\end{figure}

An important property of the nonthermal
emission is its flaring nature -- the electron ejection process is
discrete, and typical light curve of single beam is shown in
Fig.~\ref{fig_flare_structure}.  The light curve of each such flare
has a stage of fast intensity increase and sharp cut-off, its shape
reflects the properties of the magnetic field and space-time near the
horizon.

The black hole at a 100
pc distance (a sphere with this radius must contain several tens of such objects,
see  \citep{agol_2002b}) looks like a 15-25$^{\rm m}$ optical object (due
to the "thermal" spectral component) with a strongly  variable companion
in high-energy spectral bands ("nonthermal" component). The hard
emission consists of flares, the majority of which
are generated inside a 5$r_g$ distance from the BH.
These events have the durations $\sim r_g/c$ ($\sim$ 10$^{-4}$ s),
a rate of 10$^{3}$-10$^{4}$ flares per second, and an amplitude of 2\%-6\%.
The BH variable X-ray emission may be detected by modern space-borne
telescopes.

Optical emission
consists of both a quasistationary "thermal" part and a low-frequency tail
of nonthermal flaring emission. The rate and duration of optical
flares are the same as X-ray ones, while their amplitudes are
significantly smaller. Indeed, the contribution of nonthermal component
to the optical emission is
approximately $2\cdot10^{-2}$ for $\dot m = 10^{-8} - 10^{-6}$, so the mean
amplitudes of optical flares are 0.04\%-0.12\%, while the peak ones may be 1.5-2
times higher and reach 0.2\%.  Certainly, it is nearly impossible to
detect such single flares, but their collective power reaches 18-24$^{\rm
m}$ and thus may be detected in 
observations with high time resolution
($<$ 10$^{-4}$ s) by the large optical telescopes\citep{beskin_2005}.

Of course, the variability of the BH emission is related not only to the electron
acceleration processes described here. Additional variability may be result of the
plasma oscillations of different kinds, or other types of instabilities. The time scale
of such variability may be from $r_g/c$ till $r_c/c$, i.e. from microseconds till years.

\section{Search for isolated stellar mass black holes in optical range}

As it has been already noted, the main indirect observational sign of the
isolated stellar mass black hole is the absence of any spectral features in the
emission of accreting plasma (except for interstellar absorption ones)
\citep{shvartsman_1971}. Among such objects are the white dwarfs with
featureless spectra (DC-dwarfs) and known localization (as they are usually
selected as an objects with known proper motions) \citep{sion_mccook}.
On the other hand, in the last years by means of the cross-correlation of
surveys of various wave bands (from radio to gamma) and the follow-up
spectroscopic observations, the population of objects with featureless optical
spectra and without known localizations, has been detected
\citep{pustilnik_1977,tsarevsky}. The latter include the radio objects
with continuous optical spectra (ROCOSes) \citep{pustilnik_1977}. For them,
as well as for some DC-dwarfs \citep{beskin_1991}, the upper limits
on the fast optical variability as an observational manifestation of isolated
stellar mass black holes have been derived for the first
time \citep{shvartsman_1989a,shvartsman_1989b}. This idea became
the basis of the observational programme for search for BH on Russian 6-m
telescope - MANIA (Multichannel Analysis of Nanosecond Intensity Alterations)
\citep{shvartsman_1977,beskin_1997}.

Recently, some evidences appeared that the isolated stellar mass black holes
may be among the unidentified gamma-ray sources
\citep{gehrels,la_palombara_2004}.
The most promising cases are such where the independent estimation of the
object mass is possible. One of such cases is the black hole in a binary system
with a white dwarf. Such a binary may be detected by means of its periodic
brightness amplification on the tens of seconds time scale due to gravitational
microlensing \citep{beskin_2002}. By using the technical equipment of SDSS
(23$^{\rm m}$ limit in 6 square degree field) roughly 15 such objects may be
detected in 5 years. It is clear that in such system it is easy to estimate the
mass of the black hole, while there is no accretion from the white dwarf onto
it, and the black hole interacts with the interstellar medium only.
Another case when we may estimate the mass of the black hole are the long MACHO
microlensing events \citep{bennett_2002}.

Below we present the results of observations of two candidates to the isolated
stellar mass black holes.




\section{Observations of objects-candidates to isolated stellar-mass black
  holes at 6-m telescope}

Sample observations of the longest microlensing event MACHO
1999-BLG-22\citep{bennett_2002}, a most solid single
stellar-mass black hole candidate, have been performed at the Special
Astrophysical Observatory of RAS in 2003-2006
in the framework of the MANIA
experiment\citep{beskin_1997}.

\begin{figure}
  {\centering \resizebox*{1\columnwidth}{!}{\includegraphics[angle=0]{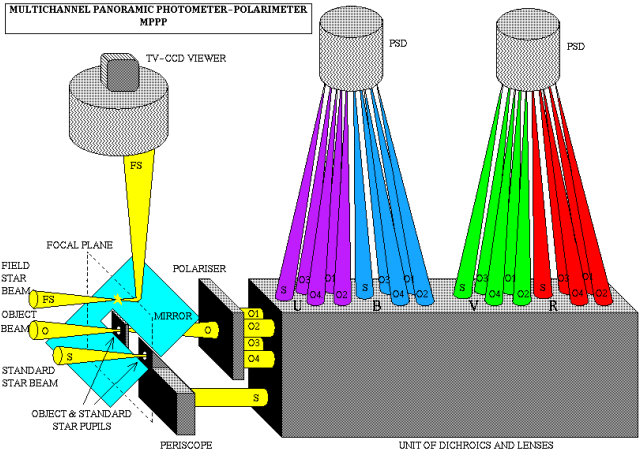} \par}}
  \caption{Optical scheme of the multichannel panoramic spectro-polarimeter (MPSP).}
  \label{fig_mpsp}
\end{figure}

We used in observations
the multichannel panoramic spectro-polarimeter (MPSP) based on position-sensitive detector
(PSD) with 1 $\mu$s time resolution \citep{debur, plokhotnichenko} (see
Fig.~\ref{fig_mpsp}).
Such detectors
use the set of multichannel plates (MCP) for electron multiplication, and multi-electrode
collector to determine its position. PSD used in our observations has the
following parameters: quantum efficiency of 10\% in the 3700-7500 A range, (S20
photocathode), MCP stack gain of $10^6$, spatial resolution of 70 $\mu$m
($0.21''$ for the 6-m telescope), 700 ns time resolution, $7\cdot10^4$ pixels
with 22 mm working diameter, and the 200-500 counts/s detector noise. The
acquisition system used is the ``Quantochron 4-480'' spectal time-code
convertor with 30 ns time resolution and $10^6$ counts/s maximal count rate.

\begin{figure}
  {\centering \resizebox*{1\columnwidth}{!}{\includegraphics[angle=0]{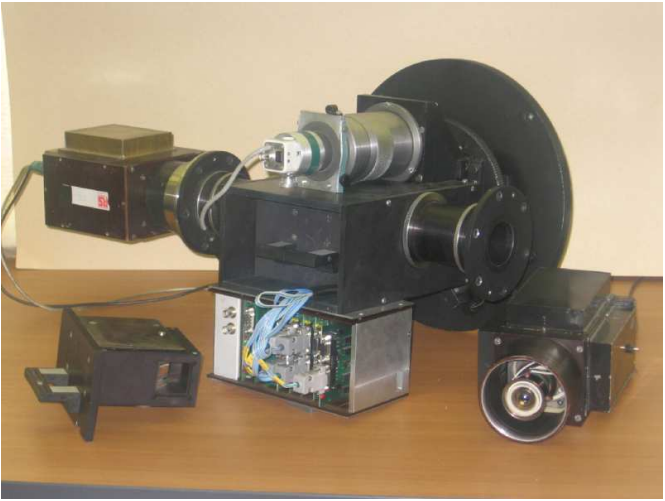} \par}}
  \caption{Multichannel panoramic spectro-polarimeter.}
  \label{fig_mpsp_image}
\end{figure}

\begin{figure}
{\centering \resizebox*{1\columnwidth}{!}{\includegraphics[angle=270]{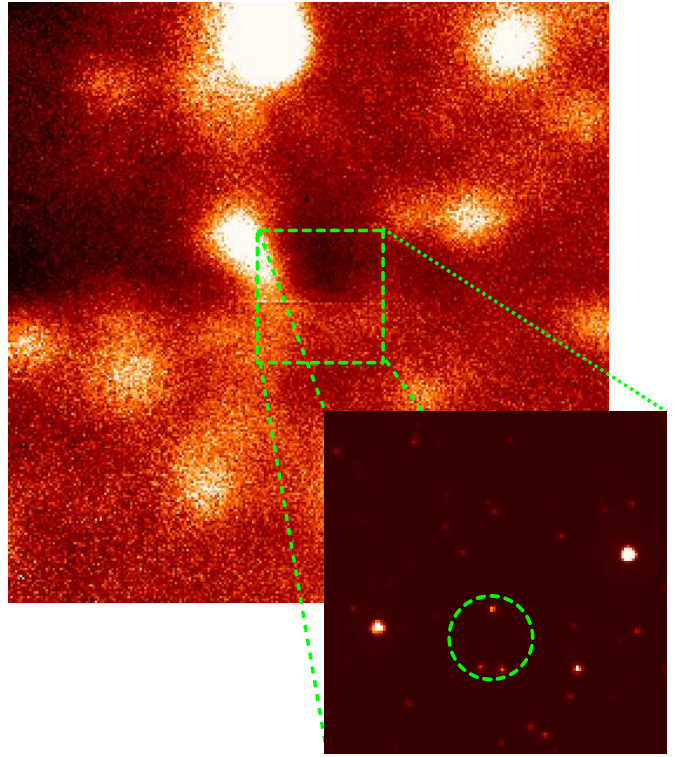} \par}}
  \caption{MACHO-1999-BLG-22, observed with panoramic photometer with high temporal resolution of Russian 6-m telescope.
    Overplotted is the Hubble ACS image with the blend resolved. Green circle is the position
    of the lens error box.}
  \label{fig_macho}
\end{figure}

\begin{figure}
{\centering \resizebox*{1\columnwidth}{!}{\includegraphics[angle=270]{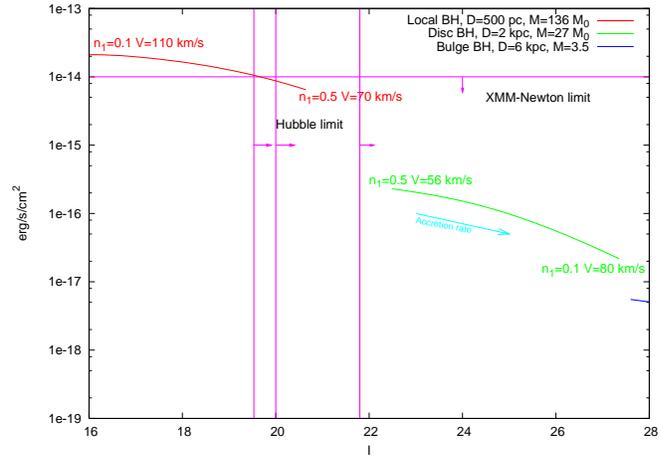} \par}}
  \caption{The prediction for the observational parameters of the MACHO-1999-BLG-22. The x-ray 
    flux is in the 0.2-10 keV band.  Overplotted are the limits from the
    XMM-Newton and Hubble ACS archival observations.}
  \label{fig_macho_limits}
\end{figure}

\begin{figure}
{\centering \resizebox*{1\columnwidth}{!}{\includegraphics[angle=270]{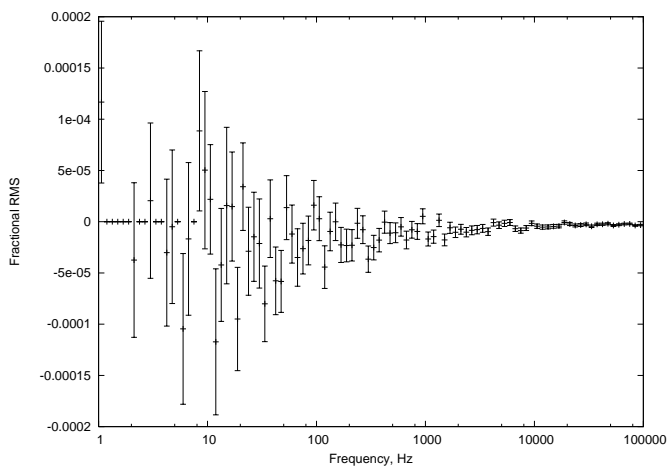} \par}}
  \caption{The optical Fourier power spectrum for the
    radio-loud x-ray source with featureless optical spectrum J1942+10. The
    upper limit for the variable emission component is 10\% in $10^{-5}$ - 1 s range.}
  \label{fig_tsar}
\end{figure}

We combined our data with the archival observations of this region
with XMM-Newton and Hubble ACS. We acquired the upper limits on
the black hole x-ray emission on the level of $10^{-14}$ erg/s/cm$^2$.

The Hubble ACS optical data allow to de-blend the original source detected by
\citep{bennett_2002} on the level of $I=19.1^{\rm m}$. We performed the
$I$-band photometry of the three stars of the blend and placed an upper limits
of $I=19.5^{\rm m}$, $I=20.0^{\rm m}$ and $I=21.8^{\rm m}$ on the $I$-band
emission of the BH for the three possible cases.

The parameters derived from the microlensing light curve don't allow
to get the unique value for the distance (and so, the mass) to the
black hole. Instead, it points to the two possible configurations of
the microlensing system, one with the lens at 500 pc (with $M=136
M_{\odot}$) and the second with 2 kpc distance ($M=3.5 M_{\odot}$). We
also studied the intermediate case (2 kpc distance, $M=27 M_{\odot}$).

The predicted observational parameters for these cases, computed
according to our model, are shown in Fig.~\ref{fig_macho_limits}, with
the observational limits overplotted.  

Our results rule out the case of the near and massive black hole.

The analysis of data from 6-m telescope do not reveal any variable source. The upper
limit on the variable emission component is placed on the $B=22^{\rm m}$ level for
the $10^{-6}$ - 100 s time scale.

The J1942+10 is the radio-loud x-ray object with featureless optical spectrum
with $B\sim18^{\rm m}$. We observed it for 40 minutes in photometric mode and
placed the upper limits on the B-band variability on the 10\% level over the
$10^{-5}$ - 1 s time scale (see Fig.~\ref{fig_tsar}). 



\section{Acknowledgements}
This work has been supported by the Russian Foundation for Basic Research
(grants No. 01-02-17857 and 04-02-17555), INTAS (grant No 04-78-7366),
and by the
grant of the Program "Origin and Evolution of Stars and Galaxies" of the
Presidium of RAS.  S.K. thanks the Russian Science Support Foundation for
support.


\end{document}